\documentclass[12pt]{article} 
\usepackage[a4paper]{typearea}
\typearea{0}
\usepackage{helvet}
\usepackage{graphicx}
\usepackage{amsmath}
\usepackage{times}
\usepackage {graphics,pstricks}
\usepackage[T1]{fontenc}
\usepackage[latin1]{inputenc}
\usepackage {pst-all}
\usepackage {multido}
\usepackage{color}
\usepackage{natbib}

\begin{document}

\begin{center}
\Large{Long-time coherent quantum behaviour of the D-Wave machine}\\
{\large D. Drakova$^1$ and G. Doyen$^2$ }  \\
{\small
$^1$ Faculty of Chemistry, University of Sofia, Bulgaria\\
URL: nhdd@chem.uni-sofia.bg \\
$^2$ Ludwig-Maximilians Universit\"at, M\"unchen, Germany \\
\vspace{-.3cm}
URL: gerold@gerold-doyen.de}
\end{center}

\begin{abstract}
Extensive experiments have demonstrated quantum behaviour
in the long-time operation of the D-Wave quantum computer.
The decoherence time of a single flux qubit is  
reported to be on the order of nanoseconds \cite{chiorescu}, which
is much shorter than the time required to carry out
a computation on the timescale of seconds \cite{johnson,dickson}.
Previous judgements of whether the D-Wave device should
be thought of as a quantum computer have been based
on correlations of the input-output behaviour of the D-Wave machine with
a quantum model,
called simulated quantum annealing, or classical models,
called simulated annealing and classical spin dynamics \cite{boixo}. Explanations for
a factor of $10^8$ discrepancy
between the single flux qubit decoherence time and the long-time coherent quantum
behaviour of many integrated flux qubits of the D-Wave device have not been offered so far.
In our contribution we investigate a model of four qubits with one 
qubit coupled to a phonon and (optionally) to 
environmental particles of high density of states,
called gravonons. The calculations indicate that when 
no gravonons are present, the current in the qubit is flipped
at some time and adiabatic evolution is discontinued. The time dependent wave
functional becomes a non-correctable superposition of many excited states.
The results demonstrate the possibility   
of effectively suppressing the current flip and allowing for continued adiabatic evolution
when the entanglement to gravonons is included.
This adiabatic evolution is, however, a coherent evolution in high
dimensional
spacetime and cannot be understood as a solution of Schr\"odinger's
time dependent
equation in four dimensional spacetime. Compared to Schr\"odinger's time development, the evolution is
considerably slowed down, though still adiabatic. The properties of our model
reflect correctly the experimentally found behaviour of the D-Wave machine
and explain the factor of $10^8$ discrepancy between decoherence time
and quantum computation time. The observation and our explanation are 
in anology to the $10^8$ discrepancy
factor found, when comparing experimental results on adsorbate quantum diffusion rate
with predictions of Schr\"odinger's time dependent equation, which can also be resolved in a model
with the coupling to gravonons included.  
\end{abstract}

\section{Introduction} \label{intro}
The D-Wave machine is a quantum computer of the quantum annealing type \cite{rose}.
It is designed to solve classical optimization problems, namely to optimize 
a cost function of many free parameters
and problems, which can be mapped on the classical Ising model \cite{ising}. 
The way to achieve the solution is by the algorithm of quantum 
annealing: slow variation of external experimental parameters,
so that, as the adiabatic theorem says, the system develops with time 
adiabatically in its energetically ground state. At the end of the 
quantum annealing time the D-Wave presents the solution of the 
optimization problem. In this sense the D-Wave machine is different from all other attempts
to construct a quantum computer.\\ 

\noindent The expectation that quantum computation
will be faster than classical is not confirmed, but nevertheless there is sound proof 
that the D-Wave machine operates as a quantum computer. 
Numerous evidences, ranging from spectoscopic data \cite{lanting} to theoretical simulations  
\cite{boixo,lanting,wang,theoryDWave}, give sound basis to accept its quantum nature.  
This macroscopic construction, consisting of hundred to more than 500 flux qubits 
on a chip, interacting with each other and subjected to transverse magnetic fields, 
behaves coherently according to Schr\"odinger's time dependent equation 
and performs coherently quantum operations over long time of
seconds and even minutes.\\

\noindent The experimental observation is that a 
single flux qubit, the building block of the D-Wave device,
a macroscopic object with $10^6-10^9$ Cooper pairs  
and dimensions of the order of $10^4$ {\rm \AA}, ''loses coherence'' within 
nanoseconds in the photon field in 
Ramsey interference experiments \cite{chiorescu}.
(More recent experiments with single flux qubits show coherence time of the order of microseconds 
\cite{bylander,stern}, however they are not comparable with the technique used by the D-Wave.) 
A single flux qubit, consisting of a supercunducting loop (Nb or Al), intercepted 
by the Josephson junction (Al oxide), 
is a two-state system with one state the persistent current running 
clockwise and the other state with the persistent current anticlockwise. In an
external transverse magnetic field the two current states are no more degenerate. 
The illumination with short (ns) near resonant photon pulse in the radiofrequency range is followed by
oscillations between the two current states, called Ramsey fringes, 
as the qubit is left to develop free with time. 
The experimental observation is: the probability for switching between the two current states  
decays with time and reaches 
a finite value for the excitation probability larger than zero.
This observation is interpreted as ''loss of coherence'' and
the suggested causes for the damping of the amplitude of the oscillations range from 
experimental imperfections and poor control of the electromagnetic environment of the flux qubit,   
to external and internal noises of different nature and the involvement of various dissipative channels (charge noise,  
photon noise, quasiparticle excitations above the superconducting gap, radiative and dielectric loss, etc.).\\

\noindent The Boltzmann factor, with the experimental excitation energy 6.6 GHz
and the experimental temperature in the millikelvin range (25 mK), is of the order of $10^{-6}$. Hence,  
the oscillations in the Ramsey interference experiments cannot be due to thermal excitation. 
Stochastic approaches, providing quantum master equation for the rate of excitation 
of a two-level system in an environment of quantum harmonic oscillators,
yield an exponential decay of the excitation probability. 
Using the parameters from the experiments, zero limiting value of the excitation   
probability and no oscillatory behaviour with time is the result \cite{stochastics,stochastics2}. 
This is in complete contrast with the experimental observations.
In a calculation based on a quantum stochastics approach the experimental observation   
of non-zero final excitation probability can be reproduced,
however, using much higher temperature than the experimental temperature. \\

\noindent The observation in an experiment with a single flux qubit, 
where its time development is not free but occurs in the presence of 
permanent illumination with radiofrequency laser, 
shows Rabi oscillations of the population   
of the first excited state, driven by the continuous laser illumination of variable duration and 
{\em immediately} followed by the measurement with the readout pulse (ref. \cite{longobardi}, p. 56, fig. 3.12).   
(This is in contrast to the Ramsey interference experiment where radiofrequency laser 
pulses of nanosecond duration are used, followed by a time interval of free development of the flux qubit.) 
The exponential decay of the amplitude of the Rabi oscillations, achieving a constant non-zero value
at time of the order of 80 ns, is attributed to decoherence due to dynamic low-frequency   
noise, which is treated within a quantum stochastics approach and a quantum master equation.  
This approach is most often used to simulate non-specified noise due to an environment. 
However, a much higher temperature of the environment,
compared with the experimental temperature of the order of mK, 
has been used to reproduce the decay of the Rabi oscillations of a single flux qubit \cite{longobardi}.\\   

\noindent A non-oscillatory exponential decay of the occupation of the excitated 
state to zero value has indeed been observed with a single  
flux qubit (ref. \cite{longobardi}, p. 50 fig. 3.10) illuminated 
with long photon pulses as a function of the delay time after the rf pulse. 
This experimental procedure, using long time illumination, allows to establish the equilibrium occupation 
of the excitated state and to follow its time development.  
As it is expected, if Boltzmann statistics dominates the distribution over the states of the flux qubit,
the probability for occupation of the excited state decays  
exponentially without oscillations, approaching the value zero at delay time of the order of 80 ns.
This trend can be fitted to Bloch-Redfield like 
master equation for a two-level system in thermal equilibrium with the electromagnetic modes.\\ 

\noindent Decoherence theory would claim that the oscillations in the Ramsey interference experiment   
with the single flux qubit are the result of collapse in one or the other current states. But then, 
if collapse in a single flux qubit occurs on the time scale of 20 ns, how can the collapses
in an interacting ensemble of five hundred flux qubits be avoided to enable  
the coherent operation of the D-Wave quantum computer?\\ 

\noindent We must conclude that quantum stochastic approaches and decoherence theory cannot 
provide the explanation of the long coherence time of the D-Wave machine, 
being unable to describe the time development of a single flux qubit.  
What the stochastic approach provides for a single flux qubit is simply a statistical Boltzmann
behaviour at temperature significantly higher than that used in the experiments. \\ 

\noindent The problem we focus on is the discrepancy of 8 orders of magnitude in the   
coherence time of a single flux qubit and the coherent operation of the D-Wave machine 
consisting of many flux qubits. We solve the time dependent Schr\"odinger equation 
with a Hamiltonian including the entanglement of the current states to gravonons, 
existing in high dimensional spacetime.            
The quantum theory lets Copenhagen quantum mechanics emerge, without postulating collapse,
and provides a deterministic definite outcome of a ''measurement'',  
hence the name Emerging Quantum Mechanics \cite{foundphysics}. 
It has been successfully used to solve the problem of localization of quantum 
particles and their appearance as classical particles \cite{localization1,localization2}, 
the problem of quantum diffusion of atoms on solid surfaces \cite{Hdiffusion}    
in a telegraph-signal like dynamics \cite{telegraph1,telegraph2}
and wave-particle duality in matter wave diffraction \cite{GDDDpresentvolume}.\\

\noindent Within the theory of Emerging Quantum Mechanics we explain
the apparent ''short coherence time'' of 20 ns of a single flux qubit (section \ref{hamiltonian1qubit}) 
and the quantum behaviour of the D-Wave machine (section \ref{hamiltonian4qubit}) over a period 
of time 8 orders of magnitude longer than that of a single flux qubit. A tentative suggestion is made to 
understand the temperature dependence of the success probability of the D-Wave computer.
Success probability is the probability to achieve the correct final result
of the posed optimization problem. Experiment with a 16 flux qubit D-Wave machine shows
a higher success probability at temperature 20-100 mK than at lower temperature \cite{dickson},
which is in total contrast with expectations for the temperature dependence    
of the performance of quantum computers.

\section{Constructing the Hamiltonian for a single flux qubit} \label{hamiltonian1qubit}
\noindent The environmental excitations are the gravonons, massive bosons which emerge  
in the limit of weak and local gravitational interaction in high dimensional spacetime (11D)
\cite{foundphysics}. The coupling to gravonons is effective only within 
spacetime deformations called warp resonances. 
The model for the non-perturbed gravonons $\{ \mid grav_i \rangle\}$ 
is a harmonic oscillator, whereas the gravonons perturbed by matter and force fields
$\{ \mid \kappa_i \rangle\}$ are the solution of the $\gamma-\eta$ model described in ref. \cite{Hdiffusion}.\\

\noindent Gravonons are bosons and every boson can be treated
mathematically as a harmonic oscillator with an effective
parabolic potential corresponding to a generalized coordinate
describing collective motion.
In 11 dimesional spacetime this is a 10D parabolic protential with  
eigenstates in 10 spacial dimensions.
The model for the non-perturbed gravonons $\{ \mid grav_i \rangle\}$ 
is a harmonic oscillator, whereas the gravonons perturbed by matter and force fields
$\{ \mid \kappa_i \rangle\}$ are the solution of the $\gamma-\eta$ model described in \cite{Hdiffusion}.
The effect of entanglement with the matter fields 
means excitation from the non-perturbed gravonon states  
$\{ \mid grav_i \rangle\}$ in $\{ \mid \kappa_i \rangle\}$ and deexcitation 
from $\{ \mid \kappa_i \rangle\}$ into $\{ \mid grav_i \rangle\}$, 
i.e. it is a scattering induced within the gravonon continuum.\\ 

\noindent The Hamiltonian for the single flux qubit plus photon is:
\begin{eqnarray} \label{hamiltsinglequbit}
{\rm H}_{qubit}
&=&{\rm H}_o+{\rm H}_{phot}+{\rm H}_{qubit-grav} \nonumber \\
&=&\sum_{j=1}^2 \left[ E_{qubit_j}a^+_{qubit_j}a_{qubit_j} +E_{w_j}a^+_{w_j}a_{w_j}
+V_{loc}^{qubit_j}(a_{qubit_j}^+a_{w_j}+a_{w_j}^+a_{qubit_j}) \right] \nonumber \\
&+&\omega_{phot}b_{phot}^+b_{phot}+\sum_{j=1}^2\left[ \varepsilon_{grav_j}c_{grav_j}^+c_{grav_j}
+\sum_{\kappa_j} \varepsilon_{\kappa_j}c_{\kappa_j}^+c_{\kappa_j}\right]  \nonumber \\
&+&V_{phot,qubit}(a_{qubit_1}^+a_{qubit_2}+a_{qubit_2}^+a_{qubit_1})(b_{phot}^++b_{phot})  \nonumber \\ 
&+&\sum_{j=1}^2 \sum_{\kappa_j}\left[ W_{grav_j,w_j}n_{w_j}c_{grav_j}^+c_{\kappa_j}+W_{w_j,grav_j}n_{w_j} c_{\kappa_j}^+c_{grav_j}\right].
\end{eqnarray}
The meaning of the symbols is $w_j$: regions in the Josephson 
junction (warp resonances), where Cooper pairs couple to gravonons;
$a_{w_j}^+, \;a_{w_j}, \;n_{w_j}$: creation, annihilation and number operator 
for persistent current in the state $\mid w_j \rangle$ in the Josephson junction;
$a_{qubit_j}^+, \;a_{qubit_j}$: creation and annihilation operator 
for persistent current in the state $\mid qubit_j \rangle $ in the flux qubit;
$b_{phot}^+, \;b_{phot}$: creation and annihilation operator for the photon;
$c_{grav_j}^+, \;c_{grav_j}$: creation and annihilation operator 
for the lowest energy gravonon in the potential, which is not perturbed 
by interaction between the matter field and the photon field;
$c_{\kappa_j}^+, \;c_{\kappa_j}$: creation and annihilation operator 
for gravonons in the deformed manifold.  \\

\noindent The terms in the Hamiltonian eq. (\ref{hamiltsinglequbit}) describe:
\begin{itemize}
\item the two current states in the qubit (first term in the second line of eq. \ref{hamiltsinglequbit}),
\item in regions of the Josephson junction where the Cooper pairs couple to the 
gravonons $w_j$ (second term in the second line of eq. \ref{hamiltsinglequbit})
\item and their interaction $V_{loc}^{qubit_j}$ (third term in the second line of eq. \ref{hamiltsinglequbit}). 
\item Term, describing the photon field (first term in the third line of eq. \ref{hamiltsinglequbit}),
\item the gravonons in the initial wave packet (second term in the third line of 
eq. \ref{hamiltsinglequbit}), as well as excited gravonons  
due to interactions with the supercurrents in the Josephson junction 
(third term in the third line of eq. \ref{hamiltsinglequbit}).
\item The interaction of the supercurrent with the photon is the standard
coupling between a photon and matter, i.e. the common dipole coupling term
(fourth line of eq. \ref{hamiltsinglequbit}).
\item The last term in eq. (\ref{hamiltsinglequbit}) is the coupling between 
gravonons and the supercurrents of the quadrupole coupling kind. 
\end{itemize}
The time dependent Schr\"odinger equation is solved for the time development
of the total wave functional: 
\begin{equation}
{\rm i}\hbar \frac{\partial}{\partial t}\Psi(t)={\rm H}_{qubit}\Psi(t), 
\end{equation}
which is represented in the basis of field configurations.
These are tensor products of the field states of the supercurrent, the photon field
and the gravonon field. In our theory the interaction with the environment is included in the 
Hamiltonian, {\em the environment is accounted for, when we solve the time 
dependent Schr\"odinger equation}. 
In contrast, stochastic approaches, starting from the Liouville equation for the 
dynamics of the density matrix with the total Hamiltonian of the system plus environment,
revert to an equation where the system alone is included in the Hamiltonian, whereas
the effects of the environment are in the Lindblad term \cite{lindblad}.\\

\noindent The fields in the system are
the persistent current of the qubit $\phi_{qubit}$, the photon field $\phi_{phot}$, the gravonon field 
comprising $\{ \mid grav \rangle \}$, $\{ \mid \kappa \rangle \}$ and $\{ \mid \lambda \rangle \}$.
The notation for the field configurations is $\mid \phi_{qubit},\phi_{phot},\kappa,\lambda\rangle$,  
for instance $\mid 1,0,0,0\rangle$ for the initial wave packet, $\mid -1,1,0,0\rangle$
for the switched current configuration plus photon, etc.
According to the assumed principle that only these warped field configurations which   
arise from the flat field configurations (non-warped configurations) 
via first order transitions are taken into account, only  
configurations of the kind $\mid \phi_{qubit},\phi_{phot},\kappa,0\rangle$ and
$\mid \phi_{qubit},\phi_{phot},0,\lambda\rangle$ are involved.\\

\noindent The physical nature of the gravonons as massive bosons 
arising from gravitons, as well as the derivation 
of an effective Schr\"odinger equation in high dimensional spacetime 
{\em yielding gravonons similar to the common quantum 
particles as its solution}, are described in detail in ref. \cite{foundphysics}.
The gravonons are the only quanta which reside not only
in four dimensional spacetime, but in the additional compactified hidden spacial 
dimensions. The decisive features of the gravonons 
are the high density of the gravonon quanta and weak coupling to the matter fields.\\
\begin{figure}{} \begin{center} \begin{minipage}{15cm}
\hspace{-.5cm}\scalebox{0.35}{\includegraphics*{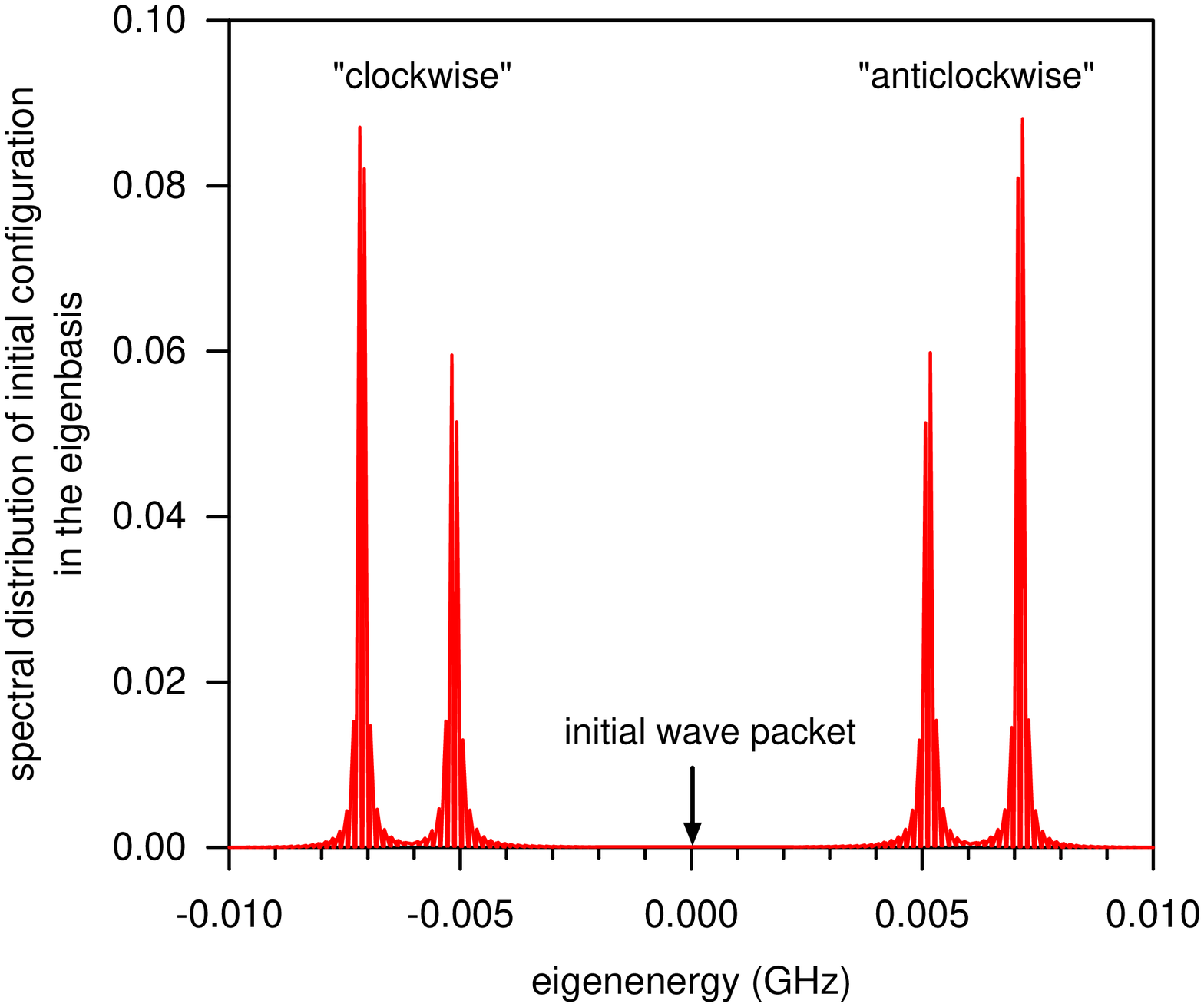}}
\end{minipage} 
\end{center}
\begin{center} \begin{minipage}{15cm}
\vspace{-6.7cm}
\hspace{7.5cm}\scalebox{0.34}{\includegraphics*{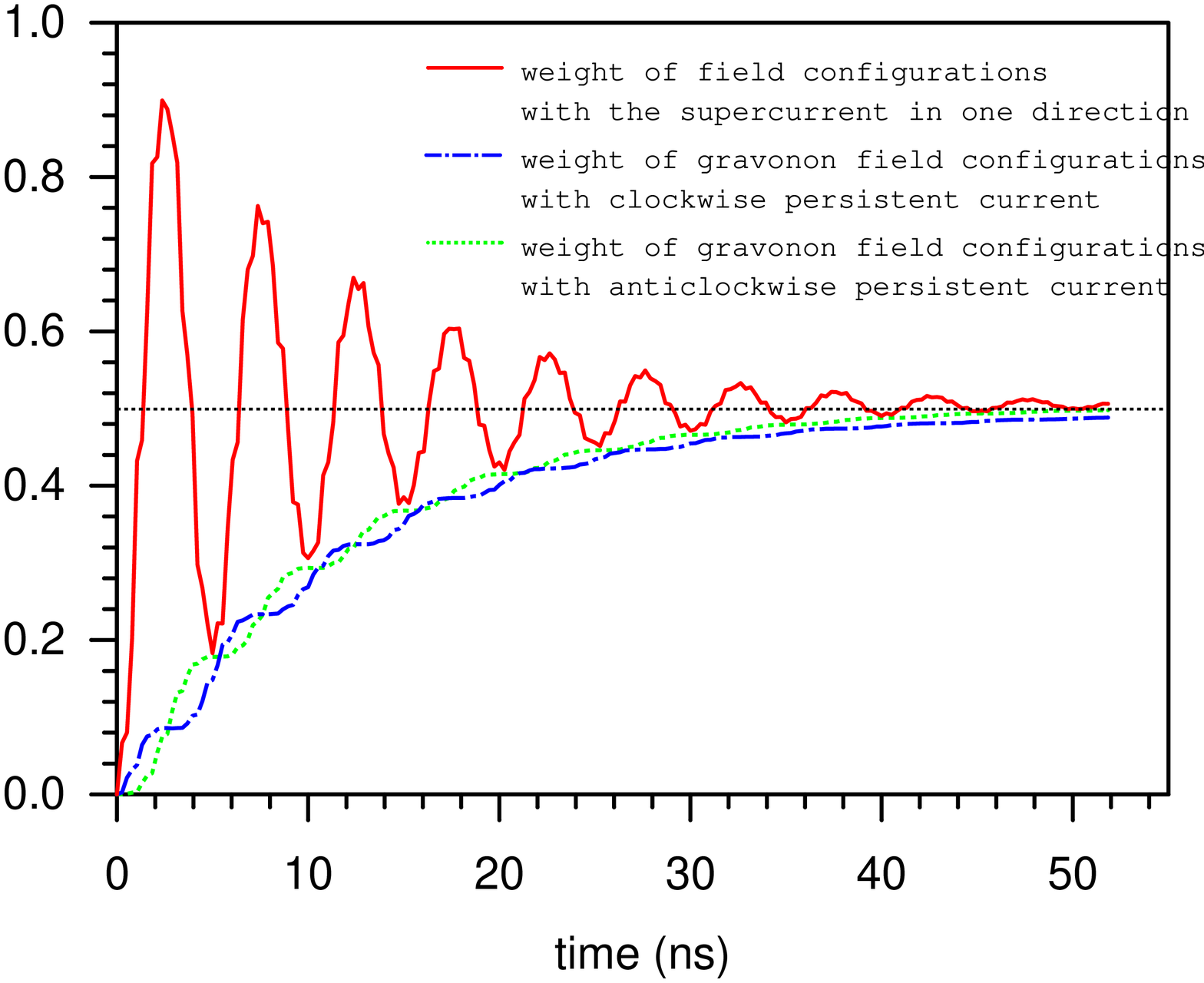}}
\end{minipage} 
\end{center}
\caption{Left panel: spectral weight of the initial wave packet $\mid 1,0,0,0 \rangle$
in the gravonon band resulting from the interactions in the local three dimensional subspace 
and the broadening due to entanglement of all local configurations with the gravonons.
The configurations at energy lower than that of the initial wave packet are mostly
with clockwise persistent current $\mid 1,0,0,0\rangle$ and $\{\mid 2,0,\kappa,0\rangle\}$. 
Those with higher energy are dominated   
by configurations with anticlockwise persistent current
$\mid -1,1,0,0\rangle$ in the flux qubit and $\{\mid -2,1,0,\lambda\rangle\}$ in the warp resonance.
\noindent Right panel: The development with time of the weight of all field configurations with the supercurrent
in one direction in a single flux qubit (full red curve) shows oscillations at  
short time which are exponentially damped, leading to a finite final value  
at long time. The reason for the damping of the oscillations is the   
development of the entanglement of the current states with gravonons which
hinders and finally precludes the current switching. The weight of field
configurations with gravonon components for the two current states 
are plotted as functions of time with green (short dashed) and blue (long dashed) curves.
\label{spectrdistr1000}} \end{figure}

\noindent Why do we need gravonons at all energies for the 
single flux qubit-plus-photon calculation and not in the case of matter wave diffraction 
experiment (cf. ref. \cite{GDDDpresentvolume})?
In the single flux qubit-plus-photon calculation we need gravonon bands at 
different energies (off-the-shell of the initial configuration) because due to the  
local interactions the local configurations are split into states with higher and lower energy 
relative to that of the initial field configuration, and the states in the 
three dimensional subsystem are no more degenerate (left panel in fig. \ref{spectrdistr1000}). 

\section{Explaining the coherence time of a single flux qubit in the photon field} 
The solution of the time dependent   
Schr\"odinger equation gives the time dependent total wave functional as a 
superposition of the basis configurations, varying with time. The initial wave packet 
in interaction with the rest configurations, but not with the gravonons, 
splits into two components  
below and above its energy, corresponding mostly to the two  
current configurations, clockwise and anticlockwise. With the entanglement with  
the gravonons they acquire finite broadening and shifts
as it is shown in the left panel of fig. \ref{spectrdistr1000}.
Each group of eigenstates has entangled to the nearly degenerate    
part of the gravonon continuum.
Oscillations between the eigenstates below and above the energy 
of the initial wave packet occur in the absence of entanglement with the gravonons.  
This is the situation for short times. 
Since the total wave functional has not yet acquired the 
components with the excited gravonons, it can be written as:
\begin{equation} \label{wavefnctnograv}
\Psi_1(t)=a(t) \mid 1,0,0,0\rangle +b(t) \mid 2,0,0,0\rangle    
+c(t) \mid -1,1,0,0\rangle +d(t) \mid -2,1,0,0\rangle    
\end{equation}
However, the entanglement with the gravonons develops with time. 
The weights of warped field configurations with the persistent current clockwise,   
as well as those with the persistent current in the anticlockwise direction, 
increase with time: 
\begin{equation} \label{wavefnctgrav}
\Psi(t)=A(t)\Psi_1(t)    
+\sum_{\kappa} k_{\kappa}(t) \mid 2,0,\kappa,0\rangle 
+\sum_{\lambda} l_{\lambda}(t) \mid -2,1,0,\lambda\rangle
\end{equation}
At the start of the time development just the terms in eq. (\ref{wavefnctnograv}) 
dominate the total wave functional
which explains the oscillations between the two current configurations.
At $t=50$ ns, when the entanglement to the gravonons  
has fully developed, only the warped field configurations in the 
total wave functional eq. (\ref{wavefnctgrav}) contribute, i.e. $A(t)$ tends to zero. 
They saturate at some final and finite values, i.e. the coefficients in the expansion of  
$\Psi(t)$ in the basis field configurations do not change with time any more.
At 50 ns current clockwise and current anticlockwise are completely entangled 
with the gravonons.
While the field configurations with excited gravonon components dominate the total wave functional 
the persistent currents cannot switch direction. 
The reason not to change with time is that, due to entanglement with the gravonons, the components 
and their contributions to the total superimposed quantum configuration cannot change while
the gravonons are in the hidden spacial dimensions. Recurrence of the gravonons back in    
three dimensional space,
disentanglement from the near degenerate gravonon band the local current configuration is originally entangled with,
and entanglement with the gravonon band, nearly degenerate with 
the current configuration of opposite direction,    
are needed to change the current configuration in the flux qubit.
As these local field configurations are not energetically degenerate, the change of current requires energy.\\

\noindent At $t=50$ ns the wave functional is still a 
superposition of many basis field configurations, but the expansion 
coefficients do not change with time. Finally we have a Schr\"odinger cat configuration, 
a superposition of field configurations, and not a tensor product, 
which, however, does not change
with time as it can be seen in fig. \ref{spectrdistr1000} (red curve in the right panel). 
{\bf The single flux qubit is in a superposition of the two persistent current configurations.}
In the time before the freezing of the current distribution in the flux qubit occurs, 
of course, the oscillations between the two current configurations
are still discernible as it is shown by the red curve in the right panel of fig. \ref{spectrdistr1000}, 
though with damped amplitude. At time $t=50$ ns a single flux qubit is frozen
in a configuration in which the relative contributions of current clockwise
and current anticlockwise in the superposition do not change.
At that time, according to our theory, the flux qubit is maximally entangled with 
the gravonon environment in a coherent state. 
It should be stressed that in our theory we do not have a new decoherence mechanism 
due to interaction with the gravonons, the gravonons are the environment  
which warrants the coherent development of the flux qubit, even suppressing
the purturbing effect of other environmental excitations in three dimensional space,  
as it is shown in section \ref{gravononphonon}.\\

\noindent The theoretical result in the right panel of fig. \ref{spectrdistr1000} is in satisfactory agreement 
with the experimental Ramsey fringes in ref. \cite{chiorescu} (fig. 4A).\\

\noindent This result might be wrongly associated with what is called ''collapse'' 
in the Copenhagen interpretation of quantum mechanics. We call it ''choice''.   
It cannot be regarded as ''collapse'' since the total wave functional is not a tensor   
product, it is a superposition of the two current configurations. 
In the theory of Emerging Quantum Mechanics, in contrast to Copenhagen quantum mechanics,
the choice is a solution of the deterministic time dependent Schr\"odinger equation, 
whereas ''collapse'' in Copenhagen quantum mechanics is postulated as random
statistically occurring, and it is not the result of any well defined dynamics. 
At time 50 ns and beyond the flux qubit is not described by a statistical distribution
over the two current configurations, as the stochastic approach yields \cite{stochastics,stochastics2}.  
The single flux qubit is not collapsed in a classical state.
It is in a quantum superposition of the two current configurations, it is in a coherent
superposition of field configurations which, however, does not change with time.\\

\noindent The reason for the observed decay of the amplitude of current 
switching in Ramsey interferometry in a single flux qubit is not 
decoherence, which implies entanglement to environmental excitations in four dimensional spacetime  
and relies on the development of the reduced density matrix into quasi-diagonal form.   
The exponential damping of the amplitude of the Ramsey fringes with time is due  
to the development of the entanglement of the local current configurations+photon with the 
gravonons leading to a coherent quantum configuration which, though it represents a quantum superimposition   
of the two current configurations, cannot change with time as long as the entanglement with the 
gravonons is effective. In this case a reduced density matrix (which is not needed in the  
theory of Emerging Quantum Mechanics) need
not acquire the diagonal form, required by decoherence 
theory and interpreted with Born's probabilistic concept.\\  

\noindent This is a rather astonishing result as the flux qubit is a macroscopic object which does not
''collapse'' in a classical state but retains its quantum behaviour for all the time preceding recurrence 
of the gravonons in three dimensional space. But it is not a unexpected result. 
How can otherwise the D-Wave machine operate as a quantum computer, if the single 
flux qubits it consists of lose coherence and collapse in classical states in the course
of the operation?\\

\section{Explaining the coherence time of the D-Wave machine: four flux qubits in transverse magnetic fields} 
\label{hamiltonian4qubit}
\begin{figure}{} \begin{center} \begin{minipage}{15cm}
\scalebox{0.45}{\includegraphics*{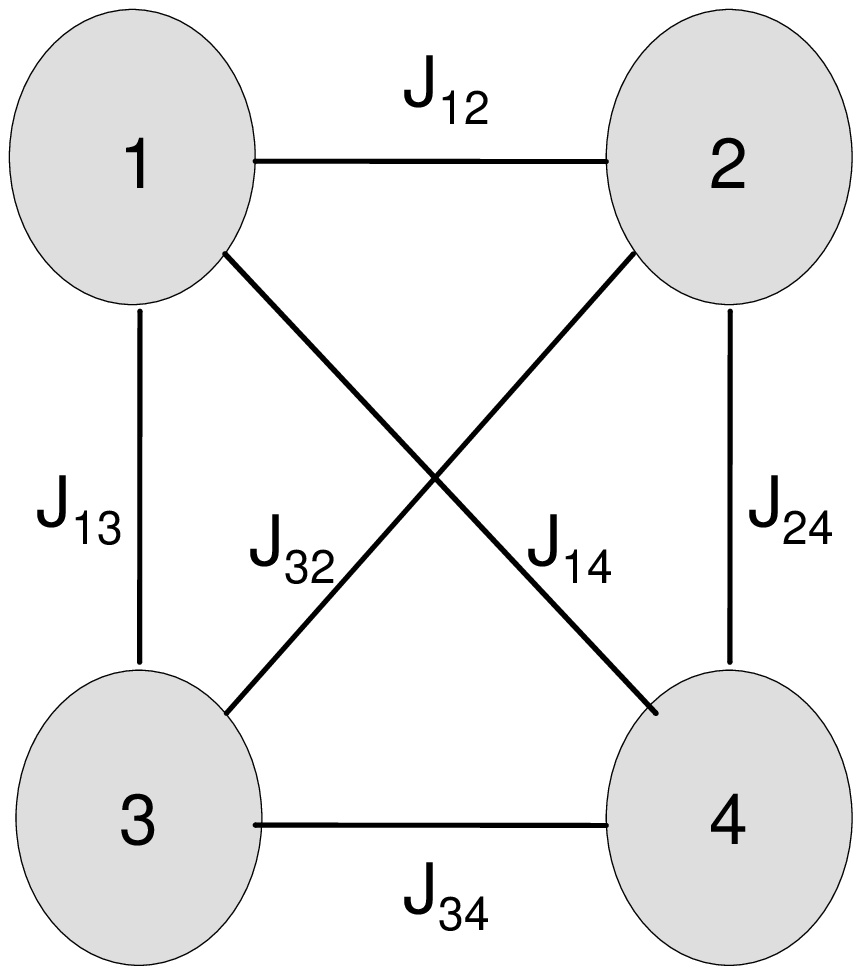}}
\end{minipage} 
\end{center}
\begin{center} \begin{minipage}{15cm}
\vspace{-5cm}
\hspace{6cm}\scalebox{0.45}{\includegraphics*{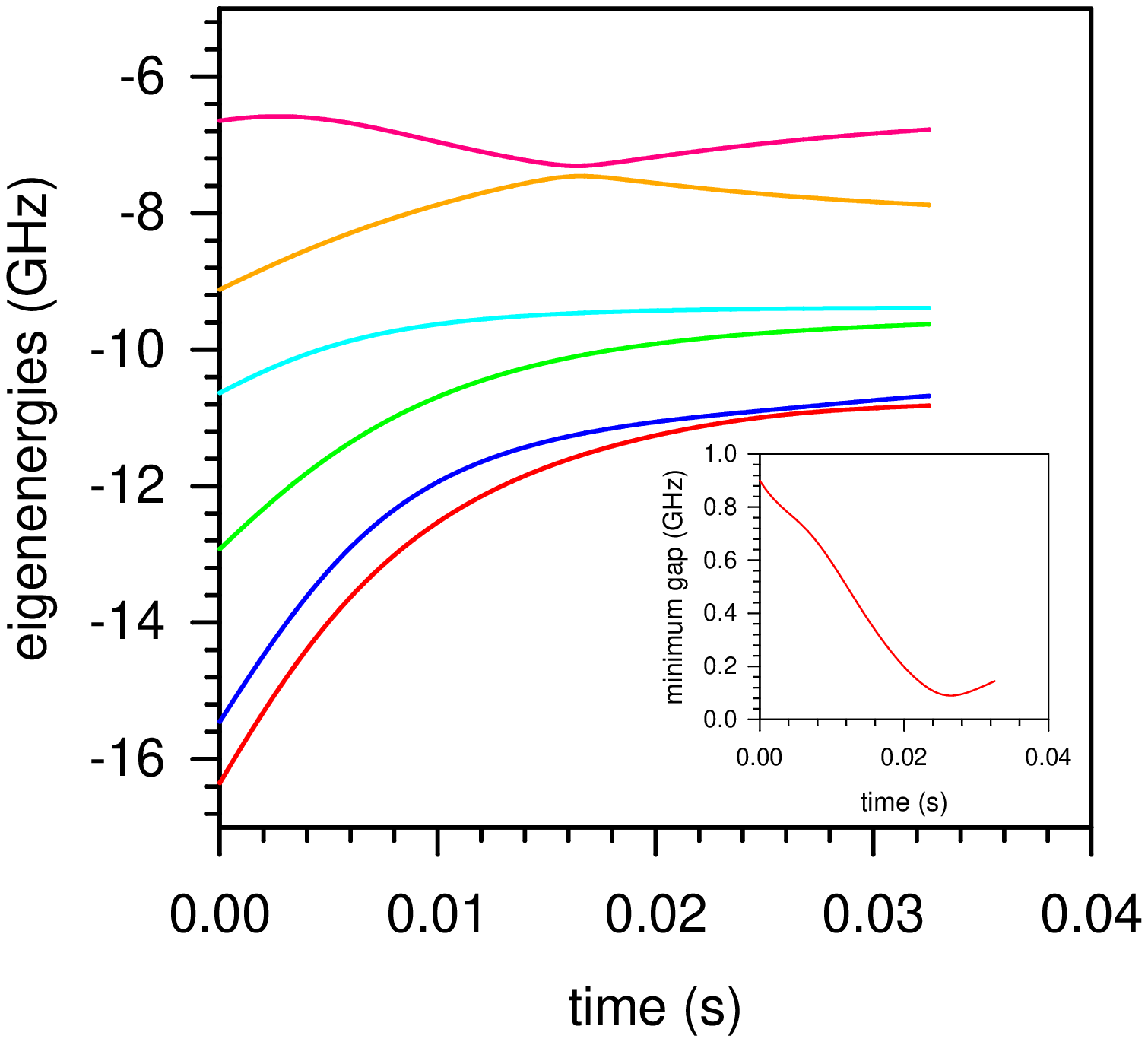}}
\end{minipage} 
\end{center}
\caption{Left panel: a model of a four flux qubit processor implementing the time dependent Hamiltonian    
eqs. (\ref{hamilt4qubits}-\ref{hamilt-vt}).
The theoretical model of the D-Wave machine consists of 4 flux qubits interacting 
with each other with interaction strengths $J_{\alpha \beta}$ and subjected
to external transverse magnetic fields with strength $ht_{\alpha}(t)$ varying with time.
Right panel: variation with time of the lowest eigenenergies of the 4 flux qubit system. 
The gap between the global energy minimum and the first excited field  
configuration is plotted in the inset, achieving a minimum value of less than 0.1 GHz at the time of avoided crossing. 
\label{model4qubits}} \end{figure}

The theoretical model of the D-Wave machine consists of four flux qubits as it is shown 
in the left panel of fig. \ref{model4qubits},
interacting with one another and subjected to external transverse magnetic fields which vary with time. 
The Hamiltonian for quantum annealing ${\rm H}_{QA}$ is now time dependent since the transverse
external magnetic fields on the flux qubits vary with time:
\begin{equation} \label{hamilt4qubits}
{\rm H}_{QA}(t) = \sum_{\alpha=1}^4 {\rm H}_{o,\alpha}+{\rm H}_P+V(t)+{\rm H}_{qubit-grav,3}.
\end{equation}
Four terms describing the non-interacting flux qubits are taken from 
the Hamiltonian in eq. (\ref{hamiltsinglequbit}), however, exempting the interaction 
with the photon field in the fourth line of this equation. 
Gravonon coupling is introduced only within {\underline{one}} qubit (the last term in eq. \ref{hamilt4qubits}).
${\rm H}_{QA}$ includes terms describing the interaction between the flux qubits:
\begin{equation} \label{hamilt-hp}
{\rm H}_P =  
\sum_{\alpha \ne \beta=1}^4 J_{\alpha \beta} \sigma_{\alpha}^z \sigma_{\beta}^z
\end{equation}
and the interaction with the external magnetic fields:
\begin{equation} \label{hamilt-vt}
V(t) = \sum_{\alpha=1}^4 ht_{\alpha}(t) \sigma_{\alpha}^x
\end{equation}
$\sigma_{\alpha}^z$ and $\sigma_{\alpha}^x$ are the Pauli matrices. 
The transverse magnetic fields are the ones which can switch the 
direction of the supercurrent in a given flux qubit and drive the superconducting current
in a superposion of currents in two opposite directions. The manipulation with time of the 
magnetic fields represents the quantum annealing procedure. 
The aim is by slowly reducing the external transverse magnetic fields $\{ ht_{\alpha} \}$ and, hence, 
the contribution in the Hamiltonian of the additional term to the non-perturbed Ising Hamiltonian,
to ensure an adiabatic time development of the system towards
the ground state of the Ising Hamiltonian, which is the
solution of the optimization problem.\\

\noindent The Pauli matrices $\sigma_{\alpha}^z$ and $\sigma_{\alpha}^x$ act 
on the current states $\mid 1 \rangle$ and $\mid -1 \rangle$:
\begin{equation}
\sigma^z \mid 1 \rangle =\left( \begin{array}{cc} 1  &  0 \\ 0 & -1 \end{array} \right) 
\left( \begin{array}{c} 1 \\ 0 \end{array} \right) = \mid 1 \rangle
\end{equation}
\begin{equation}
\hspace{1cm}\sigma^z \mid -1 \rangle =\left( \begin{array}{cc} 1 & 0 \\ 0 & -1 \end{array} \right) 
\left( \begin{array}{c} 0 \\ 1 \end{array} \right) = -\mid -1 \rangle
\end{equation}
\begin{equation}
\sigma^x \mid 1 \rangle =\left( \begin{array}{cc} 0 & 1 \\ 1 & 0 \end{array} \right) 
\left( \begin{array}{c} 1 \\ 0 \end{array} \right) = \mid -1 \rangle
\end{equation}
\begin{equation}
\sigma^x \mid -1 \rangle =\left( \begin{array}{cc} 0 & 1 \\ 1 & 0 \end{array} \right) 
\left( \begin{array}{c} 0 \\ 1 \end{array} \right) = \mid 1 \rangle .
\end{equation}

\noindent The time dependent Schr\"odinger equation is solved with a time dependent Hamiltonian:
\begin{equation} \label{psit}
\Psi(t)={\rm exp}\left[ -\frac{{\rm i} \int_0^{t} {\rm d}\tau {\rm H}_{QA}(\tau) \tau}{\hbar}\right]\Psi(0).
\end{equation}

\noindent The computational basis consists of 16 flat field configurations denoted by 4 labels,
which indicate the state of each qubit by 1 for current in the clockwise direction and -1 for current in 
the anticlockwise direction:
\begin{eqnarray}
\hspace{-.5cm}&& \mid 1111 \rangle\;\; \mid 111-1 \rangle\;\; \mid 11-11 \rangle\;\; \mid 1-111 \rangle\;\; \mid -1111 \rangle\;\; 
\mid 11-1-1 \rangle\;\; \mid 1-11-1 \rangle \nonumber \\  
\hspace{-.5cm}&&\mid -111-1 \rangle\;\;\mid 1-1-11 \rangle\;\; \mid -11-11 \rangle\;\; \mid -1-111 \rangle\;\; \mid -1-1-11 \rangle\;\; \mid -1-11-1 \rangle \nonumber \\
\hspace{-.5cm}&&\mid -11-1-1 \rangle\;\;\mid 1-1-1-1 \rangle\;\; \mid -1-1-1-1 \rangle.
\end{eqnarray}
They represent eigenstates of the Ising Hamiltonian. The notation for the gravonons are omitted, being 
in all cases the ground configurations $\mid 0 0 \rangle$ with gravonons in one qubit only. 
Thus without gravonons we have a $16 \times 16$ matrix of configurations to diagonalize to get the eigenenergies of the 
4 flux qubit system. The lowest eigenenergies and their variation 
with time can be seen in the right panel of fig. \ref{model4qubits}.\\

\noindent At $t=0$ the effect of  
$ht_{\alpha}(t=0) \ne 0, \;\alpha=1,...,4$ is to mix in all computational basis configurations, leading to 
the ground configuration of the 4 flux qubit system in the presence of the transverse magnetic fields. 
At the annealing time $t=t_f$ the     
external local transverse magnetic fields are switched off, i.e. $ht_{\alpha}(t_f)= 0, \;\alpha=1,...,4$ 
and the 4 flux qubit system provides the solution of the Ising Hamiltonian eq. (\ref{hamilt4qubits}) with $V(t_f)=0$.
The development with time will be adiabatic if
$\{ ht_{\alpha}(t)\}$ change slowly with time, leaving the system in the adiabatic 
time dependent global ground configuration,  
as it is the case shown in the right panel of fig. \ref{model4qubits}. 
If $\{ ht_{\alpha}(t)\}$ change fast with time the 4 flux qubit system
may develop diabatically with time with Landau-Zener transitions in excited eigenstates destroying the adiabaticity. 
This development is avoided by suitable choice of the time variation of the external transverse   
magnetic fields in the experiment as well as in the theory.\\

\noindent The Ising states, solution of the Hamiltonian at the end of the quantum annealing,
superimpose to build the eigenstates of the Hamiltonian with the transverse magnetic fields. 
The eigenstates possess sharp energies until the entanglement with the gravonons is included.  
And according to the adiabatic theorem the 4 flux qubit system, if it starts in the ground configuration,
will develop with time in the global ground configuration. In experiment as well as in our theory an artificial    
ground configuration is prepared by the choice of the transverse magnetic fields. 
However, the entanglement
with the gravonons may eventually destroy the adiabatic time development.
Is it so indeed? With entanglement to gravonons
the eigenstates of the 4 flux qubit system broaden into resonances within the
gravonon continuum they couple to. 
As the coupling to the gravonons is very weak and local, the broadening 
is orders of magnitude smaller than the energy differences between the lowest eigenstates 
of the 4 flux qubit system even at the time of avoided crossing 
(cf. the inset in the right panel of fig. \ref{model4qubits} showing the gap 
between the ground state and the first excited state configurations). 
Hence, the energy broadening, due to entanglement 
to gravonons, cannot lead to superpositions of the eigenstates of the 4 flux qubit system,
which would destroy the adiabaticity. \\

\noindent In analogy with the single flux qubit case, while entangled with the gravonons  
each flux qubit is frozen after 50 ns in a configuration, in which the relative contributions  
of the two current states in this flux qubit does not change with time any more. 
Each flux qubit is stabilized in this superposition, which is typical
for the global ground state configuration. 
Hence, the global ground state of the D-Wave machine cannot change either. 
Once in the ground configuration, the system of many flux qubits will stay in the
ground configuration for ever. Entanglement to gravonons  
stabilizes the global ground state configuration and helps the coherent adiabatic time 
development of the D-Wave machine towards the final solution. \\

\subsection{Destroyed adiabaticity of the D-Wave machine: four flux qubits in a phonon field} \label{phononspinflip}
\noindent Thermal excitations cannot be avoided even at the millikelvin temperatures  
at which the D-Wave machine operates. It was experimentally demonstrated in ref. \cite{dickson}
for a 16 qubit D-Wave machine that the success probability is temperature dependent: at temperature in the range
20-100 mK it is higher than at lower temperature, in contrast to other attempts for quantum computing. \\

\noindent Let a phonon be excited in just one of the flux qubits, e.g. the third flux qubit in fig. \ref{model4qubits}.
Terms are added to the quantum annealing Hamiltonian ${\rm H}_{QA}$ eq. (\ref{hamilt4qubits})
for the phonon and its interaction with the supercurrent of the kind:
\begin{equation} \label{hamilt-phon}
{\rm H}_{phon}=V_{qubit,phon}\sigma_3^x(d_{phon}^++d_{phon})+\frac{1}{2}\omega_{phon}d_{phon}^+d_{phon}.
\end{equation}
The term ${\rm H}_{phon}$ is analogous to the Jaynes-Cummings Hamiltonian of a two-level atomic system, 
coupling to the phonon field with coupling strength $V_{qubit,phon}$. 
$d^+_{phon}$ and $d_{phon}$ are creation and annihilation 
operators for the phonon mode and
$\omega_{phon}$ is the phonon frequency. With the definition of ${\rm H}_{phon}$ in eq. (\ref{hamilt-phon}) 
the effect of the phonons is to account just for switching the supercurrent 
in the third flux qubit. \\ 

\begin{figure} \begin{center} 
\begin{minipage}{15cm}
\hspace{-.5cm}\scalebox{0.3}{\includegraphics*{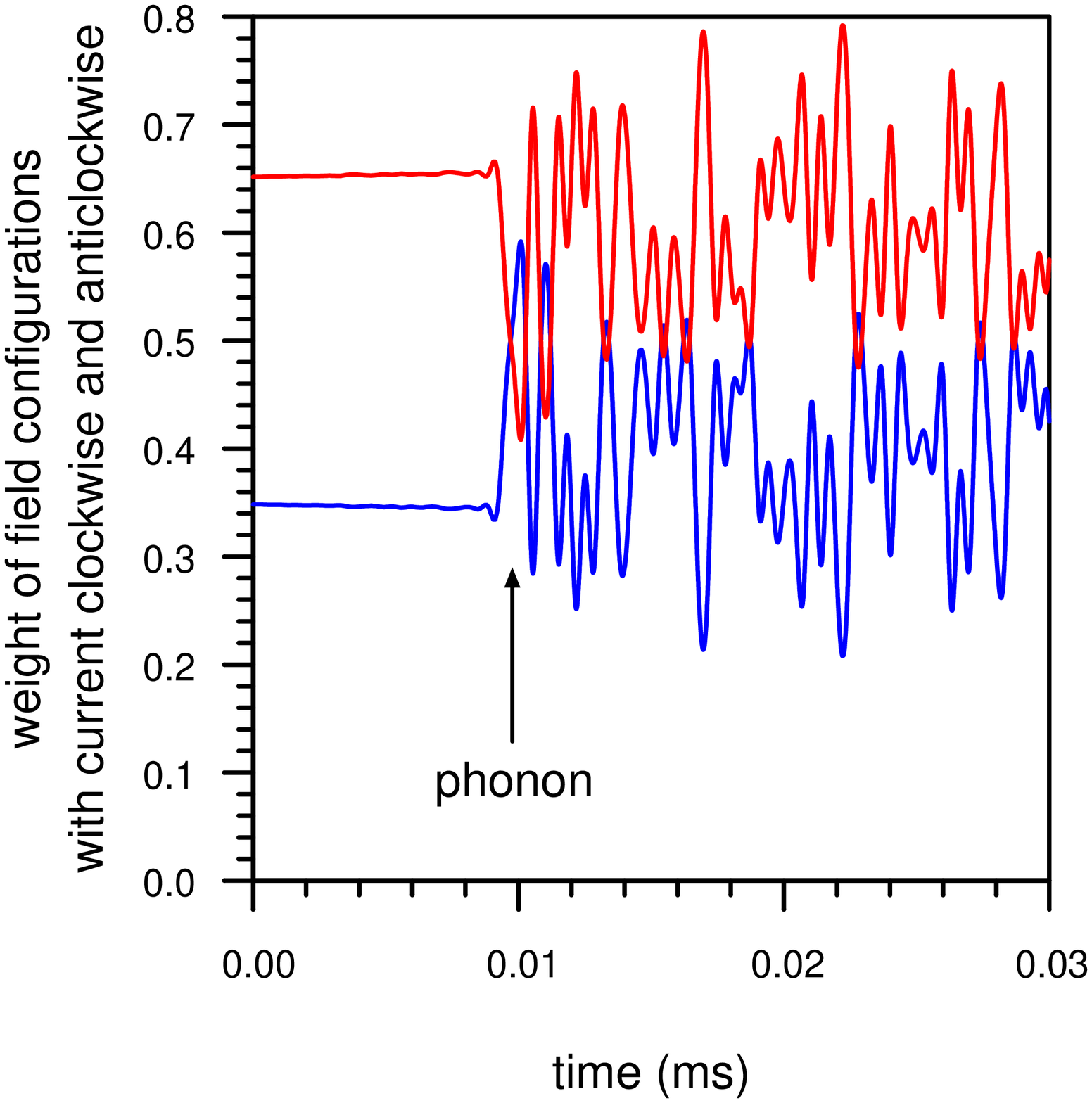}}
\end{minipage}
\begin{minipage}{15cm}
\vspace{-6.5cm}
\hspace{6.cm}\scalebox{0.35}{\includegraphics*{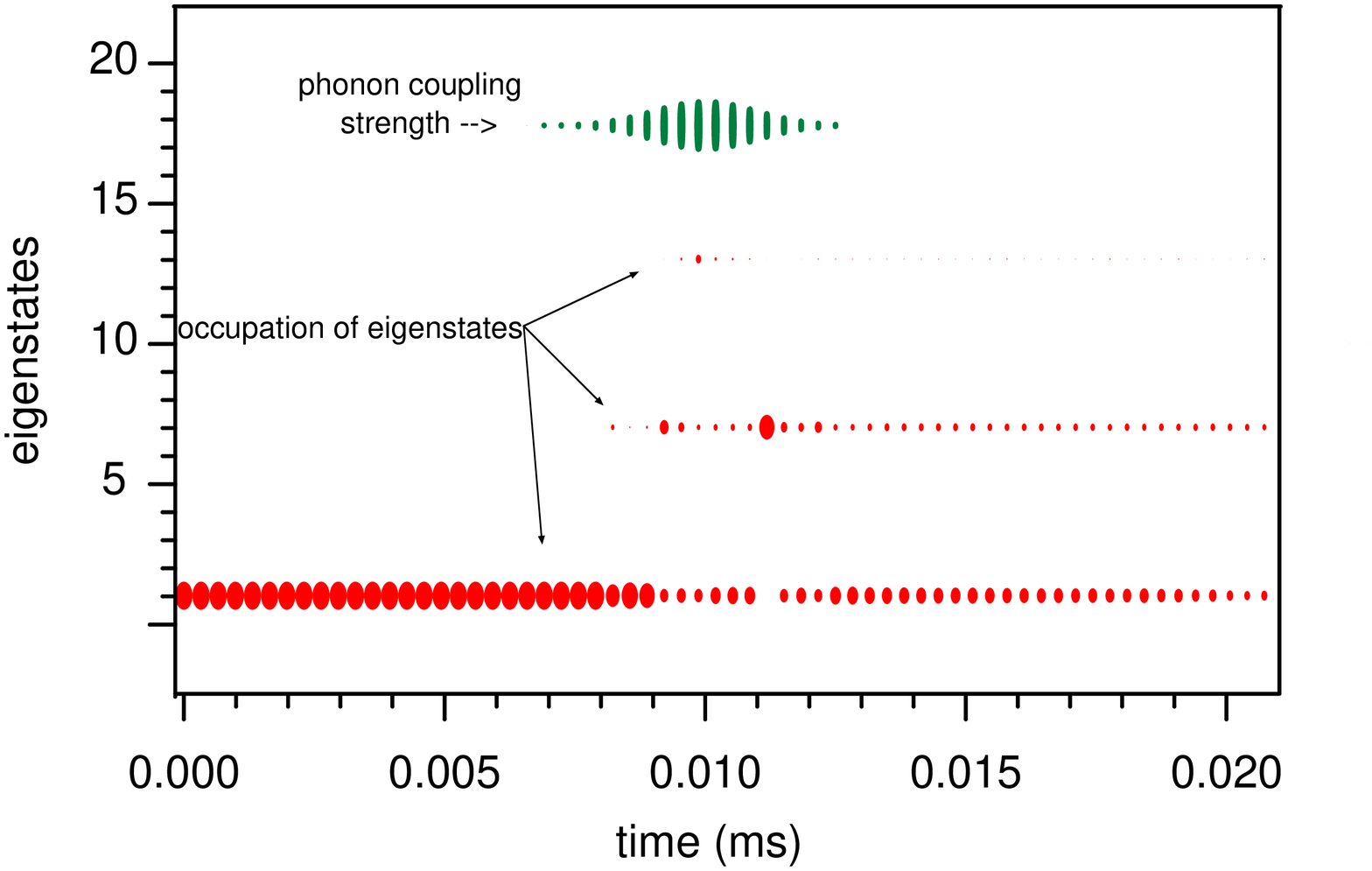}}
\end{minipage}
\end{center}
\caption{Destroyed current distribution and chaotic current flips in the third flux qubit  
as a result of the fast entanglement to the phonon field at $t\approx 0.01$ ms (left panel).  
On the right panel the destroyed adiabaticity in the global ground state configuration of the 4 flux qubit system 
as a result of the fast entanglement to the phonon field is illustrated  
with the variation of the weight of the low energy eigenstates
at the time when the coupling in the third flux qubit to the phonon is effective. 
The weight of the non-perturbed eigenstates 
which are involved in $\Psi(t)$ is displayed with varying 
size of the points. When the perturbation by the phonon is switched on
the variation of the weight of the low energy eigenstates of the 4 flux qubit system
shows a significant redistribution from the global ground state configuration over several excited configurations.
The quantity on the vertical axis is the number of the 
eigenstate of the non-perturbed 4 flux qubit system.
\label{diab-V}} \end{figure}

\noindent The result of a sudden perturbation of the 4 flux qubit system by a single phonon mode 
is destroyed adiabaticity. Figure \ref{diab-V} shows that a fast entanglement to the phonon  
field around $t=0.01$ ms perturbs the distribution of the supercurrents in the third flux    
qubit, which is typical for the global ground configuration of the 4 flux qubit system.
Dramatic and nonrepairable arbitrary switches between the currents in the flux qubit    
occur which means that the global ground configuration of the 4 flux qubit system
is changed, hence its adiabatic time development is destroyed and 
cannot be recovered. 
The coupling to the phonon leads to destroyed adiabatic time development 
in the global ground configuration of the 4 flux qubit system, as it can be seen in the plot
on the left panel in fig. \ref{diab-V}.  
The quantity on the vertical axis is the number of the 
eigenstate of the non-perturbed 4 qubit system and not its energy.
The areas of the points in the plot
scale with the weight of the eigenstates of the 4 flux qubit system, 
which get involved in the time dependent wave packet $\Psi(t)$. 
When the perturbation by the phonon is switched on 
excited field configurations of the
4 flux qubit system gain weight and a significant redistribution from the global   
ground configuration over several excited configurations occurs.
Thus, coupling to a phonon mode destroys the adiabatic time development of the system
in an irrepairable way.

\subsection{Gravonons suppress the effect of the phonon: four flux qubits in phonon and gravonon fields}
\label{gravononphonon}
\begin{figure} \begin{center} 
\begin{minipage}{15cm}
\hspace{2cm}\scalebox{0.45}{\includegraphics*{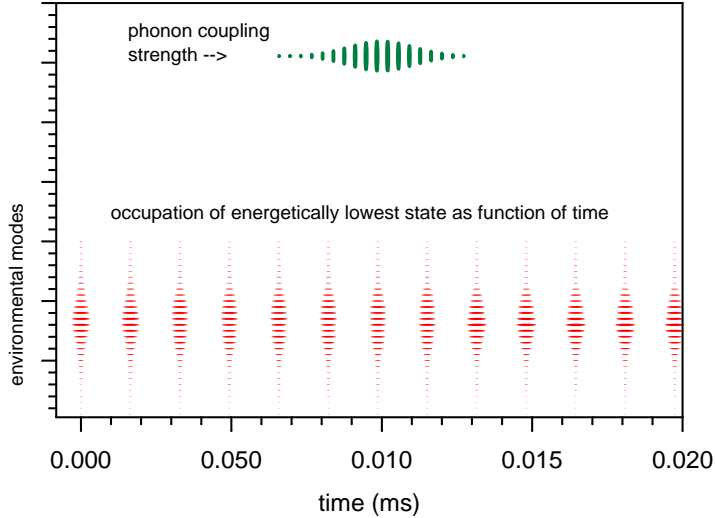}}
\end{minipage} \end{center}
\caption{Entanglement of the currents with gravonons within the Josephson junction  
suppresses the effect of the phonon and recovers the adiabaticity of the  
global ground state configuration of the 4 flux qubit system.
The quantity on the vertical axis is the number of the gravonon state 
in the gravonon continuum.
The length of the symbols in the plot scales  
with the weight of the initial wave packet in the gravonon band.
\label{adiabatic-Vph-Vgrav}} \end{figure}

\noindent In the Josephson junction the tunnelling currents 
entangle with gravonons in high spacial dimensions. The result is  
suppression of the effect of the phonons, 
and retained coherence of the system in the adiabatic ground configuration, 
as it can be seen in fig. \ref{adiabatic-Vph-Vgrav}. The length of the symbols in the plot scales  
with the weight of the initial wave packet in the gravonon band,
i.e. it is the spectral distribution of the initial wave packet in the gravonon continuum.  
The entanglement with the gravonons quenches the transitions between the low lying eigenstates
of the 4 flux qubit system due to the perturbation by the phonon. The spectral distribution of the initial 
wave packet in the gravonon continuum is not affected by the excitation of the phonon 
in one of the flux qubits.
As a consequence the many-qubit ground configuration is not modified by the interaction with the phonons.
The 4 flux qubit system develops adiabatically in the global ground configuration for long time
significantly exceding the life time of the phonon.
The gravonons suppress the effect of the phonon. 
While being entangled with the gravonons in the hidden dimensions,
the current cannot switch direction. Entanglement to gravonons stabilizes
the adiabatic ground configuration of the D-Wave machine. \\ 

\noindent The noteworthy observation for the D-Wave machine is that it achieves
higher success probability at slightly higher temperature compared to
lower temperature \cite{dickson}. The temperature, in contrast to 
expectations, improves the performance of the D-Wave machine.  
Higher temperature means phonon excitations, leading to current flips. 
This experiment shows, however, that the D-Wave quantum computer functions better when
it is slightly warmed up.  \\

\noindent Our results also show that the phonons, i.e. temperature
can destroy the D-Wave coherence by flipping the current direction  
in one flux qubit (cf. section \ref{phononspinflip}). But phonon excitation means enhanced atom vibrations.    
We know from the dynamics of chemical reactions that 
atom vibrations, i.e. temperature increase, are beneficial 
for the attenuation of the effective activation barriers for chemical reactions. Similar argument
can be used to explain the effect of temperature on the effective barrier height for tunnelling 
from the initial configuration into the global energy minimum configuration of the D-Wave machine. 
The barrier heights for state tunnelling between different minima depend upon  
the atom vibrations and, hence, upon the temperature. For small temperature increase
the effective barrier heights can get lower and then the success probability will get higher 
for shorter annealing times.
Of course, the reverse effect of destroyed superconductivity as the temperature further   
increases, leads to the attenuation of the success probability.

\section{Conclusion}

\noindent The present study of a quantum computer 
consisting of numerous flux qubits uses Emerging Quantum Mechanics,  
a method based on Schr\"odinger's time dependent quantum mechanics 
which accounts for the entanglement of fields in four dimensional spacetime 
with gravonons living in high dimensional spacetime \cite{foundphysics}. 
The astonishing result is that a single flux qubit and the D-Wave computer, 
both macroscopic objects, retain coherence and behave as quantum objects according to the 
laws of quantum mechanics with long coherence time of the order of minutes. The clue to this result is  
the entanglement of the persistent currents in the flux qubits 
with the gravonons, the massive quanta of the gravitational field,
which live in high spacial dimensions. The necessary condition    
for this result is weak coupling of the local quantum fields to an environmental continuum   
of high density of states, which is satisfied only by the gravonons in high  
spacial dimensions. \\

\noindent The major results of the present study of the D-Wave quantum computer
can be summarized:
\begin{itemize}
\item The ''coherence'' time observed experimentally of a single flux qubit is explained,
obtaining acceptable agreement with experiment. The explanation is based on
entanglement with the gravonon continuum.
\item The entanglement to gravonons also explains
the coherence time of the D-Wave quantum computer.
\item The entanglement to gravonons also explains 
why phonons do not perturb the quantum annealing process of the D-Wave 
machine and it lets the temperature dependence of the success probability of the 
D-Wave quantum computer appear comprehensible.
\end{itemize}

\noindent These results as well as the experimental observations on the D-Wave quantum
computer cannot be reproduced and explained either by stochastic quantum approaches
or within conventional decoherence theory in four dimensions. This has, however, not been
explicitely demonstrated in the paper so far. \\

\noindent Emerging Quantum Mechanics    
reproduces experimental observations of very different nature, where
orders of magnitude discrepancies have been found by computations based on conventional quantum mechanics  
in three dimensional space. They include quantum diffusion of adsorbates on solid surfaces \cite{Hdiffusion},    
and double slit diffraction experiments with massive molecules \cite{GDDDpresentvolume}.  
In these cases localization of quantum particles in three dimensional space is the result of the  
entanglement with the gravonons which live in high spacetime dimensions.

\end{document}